\documentclass[journal]{IEEEtran}
\usepackage{graphicx}
\graphicspath{{figures/}}
\usepackage{balance}
\usepackage{subfigure}
\usepackage{amssymb}
\usepackage{epsfig}

\newcommand{\Var}[2]{\mbox{Var}_{#1}\left[ {#2} \right]}

\newcommand{\ie}{{\it i.e.}}
\newcommand{\eg}{{\it e.g.}}

\newcommand{\Order}[1]{\mathcal{O}\left( {#1} \right)}

\newcommand{\mbz}[0]{\mathbf{z}}

\newcommand{\nnn}[0]{\nonumber \\ }
\newcommand{\nn}[0]{\nonumber }

\title{Through-Wall Motion Tracking Using Variance-Based Radio Tomography Networks}
\author{Joey Wilson and Neal Patwari\\
Sensing and Processing Across Networks (SPAN) Lab\\
University of Utah\\
Salt Lake City, UT, USA\\
joey.wilson@utah.edu,  npatwari@ece.utah.edu}

\begin{document}

\maketitle

\begin{abstract}
This paper presents a new method for imaging, localizing, and tracking motion behind walls in real-time. The method takes advantage of the motion-induced variance of received signal strength measurements made in a wireless peer-to-peer network. Using a multipath channel model, we show that the signal strength on a wireless link is largely dependent on the power contained in multipath components that travel through space containing moving objects. A statistical model relating variance to spatial locations of movement is presented and used as a framework for the estimation of a motion image. From the motion image, the Kalman filter is applied to recursively track the coordinates of a moving target. Experimental results for a 34-node through-wall imaging and tracking system over a 780 square foot area are presented.
\end{abstract}


\section{Introduction}

This paper explores a method for tracking the location of people and objects moving behind walls, without the need for an electronic device to be carried by or attached to the target. The technology is an extension of ``radio tomographic imaging'' \cite{Wilson09a}, which is so-called because of its analogy to medical tomographic imaging methods.  We call this extension \textit{variance-based radio tomographic imaging (VRTI)}, since it uses the signal strength variance caused by moving objects within a wireless network.  The general field of locating people or objects when they don't carry a device is also called ``device-free passive localization'' \cite{Youssef07} in contrast to technologies like active radio frequency identification (RFID) which only locate objects that carry a radio transmitter.

In a mission-critical application, we envision a building imaging scenario similar to the following.  Emergency responders, miltary forces, or police arrive at a scene where entry into a building is potentially dangerous.  They deploy radio sensors around (and potentially on top of) the building area, either by throwing or launching them, or dropping them while moving around the building.  The nodes immediately form a network and self-localize, perhaps using information about the size and shape of the building from a database (\eg, Google maps) and some known-location coordinates (\eg, using GPS).  Then, nodes begin to transmit, making signal strength measurements on links which cross the building or area of interest.  The RSS measurements of each link are transmitted back to a base station and used to estimate the positions of moving people and objects within the building.

Radio tomography provides life-saving benefits for emergency responders, police, and military personnel arriving at potentially dangerous situations. Many correctional and law enforcement officers are injured each year because they lack the ability to detect and track offenders through building walls \cite{Hunt01}.  By showing the locations of people within a building during hostage situations, building fires, or other emergencies, radio tomography can help law enforcement and emergency responders to know where they should focus their attention.

This paper explores the use of radio tomography in highly obstructed areas for the purpose of tracking moving objects through walls. First, a review of previous work and related research is summarized in Section \ref{section.related}. In Section \ref{section.VRTI}, we address a fundamentally different method for the use of RSS measurements which we call \textit{variance-based radio tomography (VRTI)}.  As the name implies, rather than use measurements of the change in mean of a link's RSS, measurements of the variance of the link's RSS values are used.  When a moving object affects the amplitude or phase of one or more multipath components over time, the phasor sum of all multipath at the receiver experiences changes, and higher RSS variance is observed. The amount of RSS variance relates to the physical location of motion, and an image representing motion is estimated using measurements from many links in the wireless network.

We briefly review the Kalman filter and apply it in Section \ref{section.kalman} to track the location of a moving object or person. In Section \ref{section.experiment}, experimental results demonstrate the use of RSS variance to locate a moving object on the inside of a building. This section also quantifies the accuracy of localization by comparing known movement paths with those estimated by the VRTI tracking system. We show that the VRTI system can track the location of an experimenter behind walls with approximately three feet average error for this experiment.

Finally, Section \ref{section.futureResearch} discusses some possibilities for future research. Advances in wireless protocols, antenna design, and physical layer modeling will bring improvements to VRTI through-wall tracking.

\section{Related Research}\label{section.related}

Previous work shows that changes in link path losses can be used to accurately estimate an image of the attenuation field, that is, a spatial plot of attenuation per unit area \cite{Wilson09a}.  Experimental tests show that in an unobstructed area surrounded by a network of nodes, the estimated image displayed the positions of people in the area.

Indoor radio channel characterization research demonstrates that objects moving near wireless communication links cause variance in RSS measurements \cite{Bultitude87}. This knowledge has been applied to detect and characterize motion of network nodes and moving objects in the network environment \cite{Woyach06}. Polarization techniques have also been used to detect motion \cite{Pratt08}. These studies focus mostly on detection and velocity characterization of movement, but do not attempt to localize the movement as the work presented in this paper does.

Youssef, Mah, and Agrawala \cite{Youssef07} demonstrated that variance of RSS on a number of WiFi links in an indoor WLAN can be used to (1) detect if motion is occuring within a wireless network, and (2) localize the moving object based on a manually trained lookup. In most emergency situations, however, manual training is not possible since it can take a significant amount of time and access to the area being tracked is restricted.

Real-time location systems (RTLS) are based on a technology that uses electronic tags for locating objects. For logistics purposes in large facilities, commercial real-time location systems are deployed by installing infrastructure in the building and attaching active radio frequency identificaiton (RFID) tags to each object to be tracked. RTLS systems are not useful in most emergency operations, however, since they require setup inside of a building prior to system use.  Further, RTLS systems cannot locate people or objects which do not have an RFID tag.  In emergencies, an operation cannot rely on an adversary wearing a tag to be located.  Thus, tag-based localization methods are insufficient for most emergency operations.

An alternate tag-free localization technology is ultra-wideband (UWB) through-wall imaging (TWI) (also called through-the-wall surveillance).  In radar-based TWI, a wideband phased array steers a beam across space and measures the delay of the reflection response, estimating a bearing and distance to each target. Through-wall radar imaging has garnered significant interest in recent years  \cite{Aryanfar04,Lin05,Lin06,Song05,Vertiy04}, for both static imaging and motion detection. Commercial products include Cambridge Consultants' Prism 200 \cite{Cambridge} and Camero Tech's Xaver800 \cite{Camero}, and are prohibitively expensive for most applications, on the order of US \$100,000 per unit. These products are accurate close to the device, but inherently suffer from accuracy and noise issues at long range due to monostatic radar losses.  In free space at distance $d$, radar systems measure power proportional to $1/d^4$, in comparison to  $1/d^2$ for radio transmission systems.

Radio tomography takes a fundamentally different approach from traditional TWI systems by using large networks of sensors. While initial attempts \cite{Hunt05} have allowed 2-4 high-complexity devices to collaborate in TWI, this research relies on tens or hundreds of collaborating nodes to simultaneously image a larger area than possible with a single through-wall radar.  RTI's imaging capability increases as $\Order{N^2}$ for $N$ sensors, thus large networks, rather than highly capable nodes, lead to improved imaging and tracking capabilities.

Multistatic radar research has also developed technologies called multiple-input multiple-output (MIMO) radar. These technologies also use distributed devices, perhaps without phase-synchronization, in order to measure radar scattering \cite{Haimovich08}.  The use of many distributed antennas is a type of spatial diversity for a radar system which can then avoid nulls in the radar cross-section (RCS) of a scattering object as a function of scattering angle \cite{Fishler06}.

MIMO radar is a complementary technology to radio tomography.  While MIMO radar measures scattering of the transmitted signal by the object of interest, radio tomography methods are based on measurements of transmission through a medium. Integration of the two modalities is beyond the scope of this paper, but is perhaps a promising direction for future research.

\section{Variance-Based Radio Tomographic Imaging} \label{section.VRTI}

\subsection{Overview}

In a multipath environment, a wireless signal travels along many paths from transmitter to receiver. Each multipath has associated with it an amplitude and phase, and the received signal is a summation of each incoming multipath component. The complex baseband voltage for a continuous-wave (CW) signal measured at a receiver can be expressed as \cite{durgin02},
\begin{equation}\label{eq.voltageReceiver}
 \tilde{V} = \nu + \sum_{i=1}^L V_i \exp\left( j\Phi_i\right),
\end{equation}
where $\nu$ is a noise component, $V_i$ is the magnitude and $\Phi_i$ is the phase of the $i$th multipath component (wave) impinging on the receiver antenna. Note that what we call the ``received signal strength'' (RSS) is actually the measurement of the received power in decibel terms $R_{dB} = 10 \log_{10} \| \tilde{V}\|^2$.

When motion occurs near a wireless link, some of the multipath components may be affected. We quantify the intuition that motion in spatial areas where many multipath exist causes more variance of the RSS. For example, objects moving near a node will usually cause larger fluctuations in RSS on a link than the same objects moving at positions far away from either node \cite{Bultitude87}. By combining RSS variance information for many links in a wireless network, motion can not only be detected, it can be localized.

In this section we describe how each link's RSS variance is dependent upon the power contained in multipath components that are affected by moving objects. We quantify this relationship for Ricean fading environments, then provide a linear model relating observed variance to the spatial location of movement.  Finally, a formulation for estimating an image of object motion in space is presented.

\subsection{Log-Ricean RSS During Movement}

Assume that multipath component $i$ travels through a subset of space $\mathcal{S}_i$.  This subset $\mathcal{S}_i$ might be some narrow volume around the line tracing its path from the transmitter to receiver.   We assume that if object motion occurs at a position $\mbz \in \mathcal{S}_i$, the phase $\Phi_i$ changes randomly.  For example, if an object moves into and occupies part of $\mathcal{S}_i$, its presence in the path causes a change due to the scattering from, transmission through, or diffraction around the object.

In particular, we assume that when $\mbz \in \mathcal{S}_i$, the distribution of $\Phi_i$ becomes uniform on $[0, 2\pi)$.  Because sizes of moving objects are typically on the order of or larger than a wavelength, and wave phase is a very sensitive function of the electromagnetic properties of objects in its path, it is reasonable to assume that the phase changes unpredictably and randomly.

Now, consider the case when there is motion of objects at several positions $\mathcal{Z}$, a set of coordinates in space.  We group the multipath into:
\begin{itemize}
\item \textit{Changing multipath}: Multipath components $i$ which are impacted by the motion, \ie, $\mbz \in \mathcal{S}_i$ for some $\mbz \in \mathcal{Z}$.
\item \textit{Static multipath}: Multipath components $i$ which are not impacted by motion, \ie, $\mbz \notin \mathcal{S}_i$ for all $\mbz \in \mathcal{Z}$.
\end{itemize}
Rearranging (\ref{eq.voltageReceiver}), 
\begin{equation}\label{eq.voltageReceiverGrouped}
\tilde{V} = \nu + \sum_{i \in \mathcal{T}} V_i \exp\left( j\Phi_i\right) + \sum_{i\notin \mathcal{T}} V_i \exp\left( j\Phi_i\right) 
\end{equation}
where $\mathcal{T} = \{i:\mbz \notin \mathcal{S}_i \forall \mbz \in \mathcal{Z}\}$ is the set of static multipath.  The first summation is of the static multipath, while the second sum is of the changing multipath. Since the first summation is non-random and does not change, we can write it simply as one voltage and phase,
\begin{equation}\label{eq.voltageReceiverGrouped2}
\tilde{V} = \nu + \bar{V} \exp\left( j\bar{\Phi} \right) + \sum_{i\notin \mathcal{T}} V_i \exp\left( j\Phi_i\right) 
\end{equation}
where $\bar{V} = \left|\sum_{i \in \mathcal{T}} V_i \exp\left( j\Phi_i\right) \right|$ and $\bar{\Phi} = \angle \sum_{i \in \mathcal{T}} V_i \exp\left( j\Phi_i\right)$.  Assuming that the multipath components outside $\mathcal{T}$ are several, we apply the central limit theorem and assume that the real and imaginary parts of the changing multipath sum are independent and identically distributed (i.i.d.) zero-mean Gaussian random variables $V_{ns,I}$ and $V_{ns,Q}$ \cite{durgin02},
\begin{equation}\label{eq.voltageReceiverRandom}
\tilde{V} = \nu + \bar{V} \exp\left( j\bar{\Phi} \right) + V_{ns,I} + j V_{ns,Q}
\end{equation}
The variances of $V_{ns,I}$ and $V_{ns,Q}$ plus the variance of noise $\sigma^2_{\nu}$ is denoted $\sigma^2$ and is the sum of the powers of the changing multipath components,
\begin{equation}\label{eq.NS_power}
\sigma^2 = \sigma^2_{\nu} + \sum_{i\notin \mathcal{T}} |V_i |^2.
\end{equation}

The envelope of $\tilde{V}$, \ie, $R = |\tilde{V}|$, is a Ricean random variable $R$ with pdf,
\begin{equation}
f_R(r) = \frac{r}{\sigma^2} \exp\left(-\frac{r^2+\bar{V}^2}{2\sigma^2} \right) I_0 \left(\frac{r \bar{V}}{\sigma^2} \right)
\end{equation} 
where $I_0(\cdot)$ is the zero-order Bessel function of the first kind. For received power $R^2$ measured in dB units, \ie, $R_{dB} = 20 \log_{10} R$, we can use the Jacobian method to show that $R_{dB}$ has the log-Ricean pdf,
\begin{eqnarray} \label{eq.logRiceanPDF}
f_{R_{dB}}(r) &=& f_R(\exp(c r )) c \exp(cr) \nn \\
&=& \frac{c e^{2 c r}}{\sigma^2} \exp\left(-\frac{e^{2 c r}+\bar{V}^2}{2\sigma^2} \right)
                I_0 \left(\frac{e^{c r} \bar{V}}{\sigma^2} \right)  \nnn
\end{eqnarray}
where $c = \frac{\log 10}{20}$.

\subsection{RSS Variance}

\begin{figure}[htbp]
\centering
\includegraphics[scale=.55]{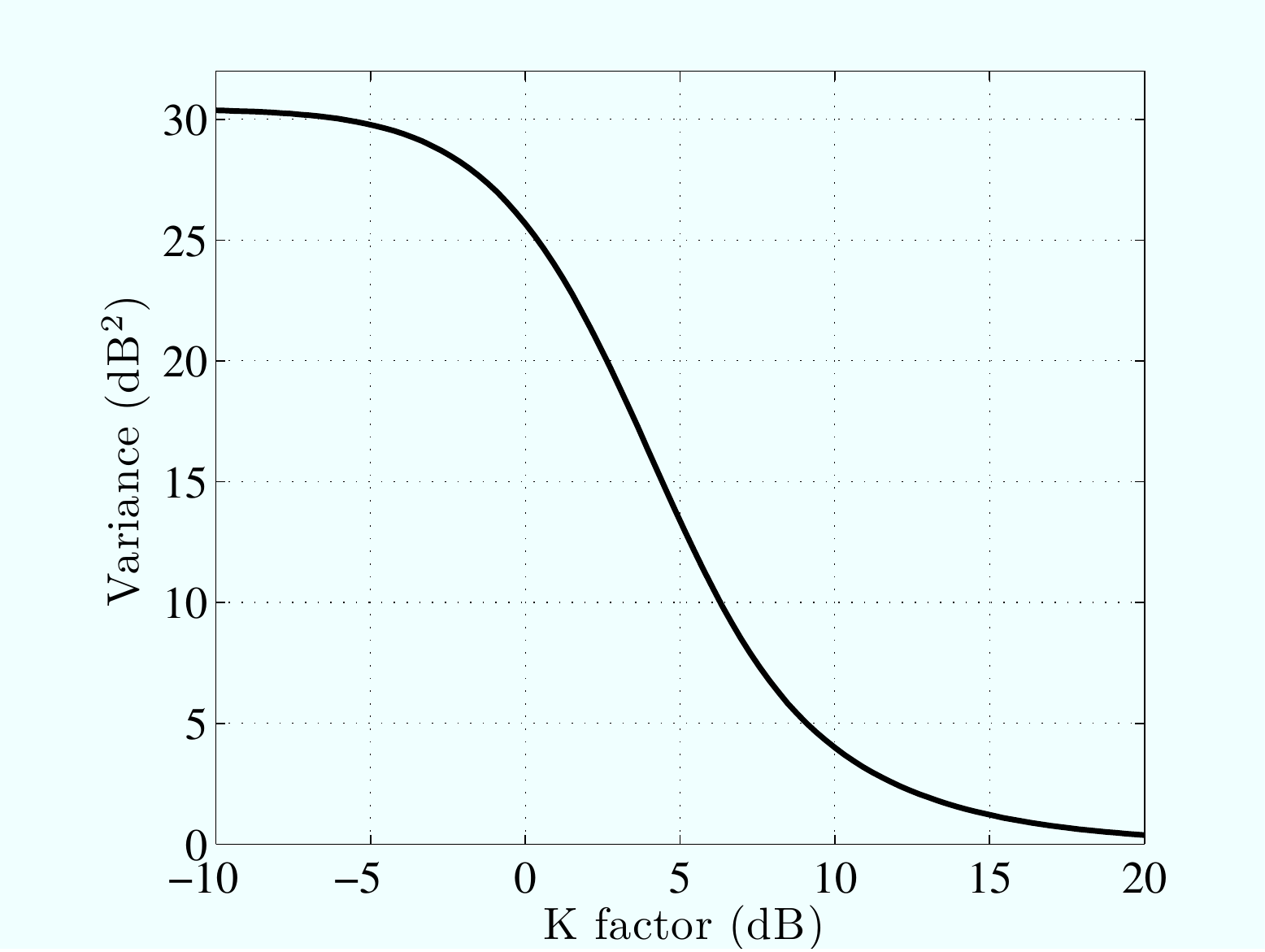}
\caption{The variance of a log-Ricean random variable vs.~$K_{dB}$.}\label{fig.plotVarianceOfLogRicean}
\end{figure}

Variance of $R_{dB}$ is defined as
\begin{eqnarray} \label{eq.varianceRSS}
\Var{}{R_{dB}} &=& \int \left(r-\mu_{R_{dB}} \right)^2 f_{R_{dB}}(r) dr , \nnn
\mu_{R_{dB}} &=& \int r f_{R_{dB}}(r) dr . 
\end{eqnarray}
The $K$ value of a Ricean pdf is defined as \cite{durgin02}
\begin{equation} \label{eq.K}
 K  = \frac{\bar{V}^2}{2\sigma^2}
\end{equation}
In dB terms, and applying (\ref{eq.NS_power}),
\begin{equation} \label{eq.K_dB}
K_{dB} = 10\log_{10} K  = - 3  + 10 \log_{10} \bar{V}^2 - 10 \log_{10} \sum_{i\notin \mathcal{T}} |V_i |^2
\end{equation}
Using the theoretical model in (\ref{eq.logRiceanPDF}), the variance of RSS (\ref{eq.varianceRSS}) is calculated using numerical integration, and is plotted in Figure \ref{fig.plotVarianceOfLogRicean} as a function of $K_{dB}$.  Note that the RSS variance (\ref{eq.varianceRSS}) is purely a function of $K$ -- for constant $K$, the scale of $\sigma^2$ and $\bar{V}^2$ do not change the variance.

We conclude from (\ref{eq.K_dB}) and Fig. \ref{fig.plotVarianceOfLogRicean} that the RSS variance due to motion in a network is largely a function of the power contained in the non-static multipath components. When no motion occurs, the K-factor is very high, since no power is contained in the changing multipath components.  When motion occurs, certain multipath components are affected, the K-factor becomes much lower, and the RSS variance increases.

\subsection{Linearity of RSS Variance to $K_{dB}$}
As seen in Fig. \ref{fig.plotVarianceOfLogRicean}, the log-Ricean random variable has a region of $K_{dB}$ where its variance is approximately linear with $K_{dB}$. This linear region is between -2 dB and +10 dB, and corresponds to the static multipath power being between 37\% less than, to 10 times greater than, the changing multipath power. The non-linearity for high and low $K$ are not of primary concern to variance-based RTI for the following reasons:
\begin{itemize}
\item $K_{dB}<-2$ dB: When $K$ is low, the changing multipath are stronger in power than the static multipath sum power.  In these cases, the moving objects are also likely to cause a noticeable reduction in the mean received power, because they typically cause some shadowing of the changing multipath.  As a result, mean-based RTI can be used instead of (or in some combination with) variance-based RTI.  Note that with low $K$, we do see high variance, so motion can be detected using variance-based RTI, but the non-linearities make the linear model, and thus the image estimator, less accurate.
\item $K_{dB}>10$ dB: This corresponds to the case that the static multipath are more than ten times stronger in power than the changing multipath.  That is, the motion of the moving objects causes changes only in a few, low-power multipath on the link.  In this case, it is difficult to distinguish the changes caused by motion from noise.  The variance of 1 dB$^2$ due to motion at a level of $K_{dB}=16$ may be difficult to distinguish from noise effects (quantization, thermal noise, etc.) on standard radio hardware.  At these high $K$ levels, it seems unlikely that variance-based RTI will be successful.
\end{itemize}

In short, the linear region of RSS variance measurements is the most important region for purposes of variance-based RTI.

\subsection{Linear Image Model}
The goal of a variance-based RTI system is to use RSS variance measurements on $M$ links in a wireless network to determine an image vector $\textbf{x}$ that describes the presence of motion occuring within $N$ voxels of a physical space. Since voxels locations are known, VRTI allows one to know where the moving objects are most likely located.

The image vector $\textbf{x}$ is a representation of motion occurring within each spatial voxel of the network area, described mathematically by
\begin{equation}
x_{j} = \left\{ {
\begin{array}{cc}
1 & \mbox{if motion occurs in voxel $j$} \\
0 & \mbox{otherwise} 
\end{array}}
\right .
\label{eq.x}
\end{equation} 

The variance on each link caused by moving objects can be approximated as a linear combination of the movement occuring in each voxel, weighted by the amount of variance that motion in that particular pixel causes on the link's RSS.
\begin{equation}
Var(R_{dB}) = \sum_j w_{j} x_j + n,
\end{equation}
where $n$ is measurement noise and modeling error, and $w_{j}$ is the variance caused by movement in voxel $j$. If all links are taken into account, the system can be expressed in matrix form as:
\begin{equation}
\textbf{s} = \textbf{Wx} + \textbf{n}
\label{eq.linearModel}
\end{equation}
where $\textbf{s}$ is an $M\times1$ vector of the variance for each link, $\textbf{W}$ is an $N \times M$ matrix representing the variance weighting for each pixel and link, $\textbf{n}$ is a $M\times1$ noise vector, and $\textbf{x}$ is the $N \times 1$ motion image to be estimated.

\subsection{Elliptical Weight Model}\label{section.Model.NEW}
If knowledge of an environment were available, one could estimate the variance weights $w_j$ for each link. Perhaps calibration measurements or ray tracing techniques could aid in estimation of the linear transformation $\textbf{W}$. For time-critical emergency operations, one cannot expect to obtain floor plans and interior arrangements of the building. With no site-specific information, we require a statistical model that describes the contribution of motion in each pixel to a link's variance.

One such statistical model has been described for link shadowing is the normalized elliptical model \cite{Agrawal08}. Consider an ellipsoid with foci at the transmitter and receiver locations. The excess path length of multipath contained within this ellipsoid must be less than or equal to a constant.  Excess path length is defined as the path length of the multipath minus the path length of the line-of-sight component. As described in previous sections, the variance of a link's RSS is highly related to the power contained in the mulipath components affected by motion. With this reasoning, we make the assumption that motion occuring on voxels within an ellipsoid will contribute significantly to a link's RSS variance, while motion in voxels outside will not. This is a binary quantization, but provides a simple, single-parameter spatial model.

The variance weight for each voxel decreases as the distance between two nodes increases.  As the link gets longer, the amount of power in the changing multipath components is decreased along with the link's RSS variance. The relationship of link distance to voxel variance weighting is a topic of future research, but emprical tests have indicated that dividing the variance weighting by the root of the link distance generates images of higher quality than other functions tested. The weighting is described mathematically as
\begin{equation}
[W]_{l,j} = \frac{1}{\sqrt{d_l}}\left\{ {
\begin{array}{cc}
\psi & \mbox{if $d_{lj}(1) + d_{lj}(2) < d_l + \lambda$} \\
0 & \mbox{otherwise} 
\end{array}}
\right .
\label{eq.ellipse}
\end{equation} 
where $d_l$ is the distance between the two nodes, $d_{lj}(1)$ and $d_{lj}(2)$ are the distances from the center of voxel $j$ to the two respective node locations on link $l$, $\psi$ is a constant scaling factor used to normalize the image, and $\lambda$ is a tunable parameter describing the excess path length included in the ellipsoid.

The normalized ellipse weight model is certainly an approximation, but experimental data has shown its effectiveness for VRTI, as will be shown in a later section. Future work will use theoretical arguments and extensive measurements to develop a statistical model of RSS variance as a function of location.

\subsection{Process Sampling, Buffering, and Variance Estimation}\label{section.VRTI.processSampling}
In this paper, it is assumed that the link signal strength process is sampled at a constant time period $T_s$, resulting in the discrete-time signal for link $l$:
\begin{equation}
R_l[k] = R_{dB_l}(kT_s).
\end{equation} 
where $R_{dB_l}(kT_s)$ is the RSS measurement in dB at time $kT_s$ for link $l$. It is also assumed that the process remains wide-sense stationary for a short period of time. These assumptions allow the recent variance of the process to be estimated from a history buffer of the previous $N_B$ samples for each link. The short-term unbiased sample variance $\hat{s}_l$ for each link $l$ is computed by
\begin{equation}
\hat{s}_l = \frac{1}{N_B-1} \sum_{p=0}^{N_B-1} (R_l[k-p] - \bar{R}_l[k])^2
\end{equation} 
where
\begin{equation}
\bar{R}_l[k] = \frac{1}{N_B} \sum_{p=0}^{N_B-1} R_l[k-p]
\end{equation}
is the mean of the signal strength buffer. The sample variance vector for all links in the wireless network is
\begin{equation}
\hat{\textbf{s}} = \left[ \hat{s}_1, \hat{s}_2, ... , \hat{s}_M \right] ^T
\end{equation} 

\subsection{Regularization and Image Estimation}
The linear model (\ref{eq.linearModel}) provides a mathematical framework relating movement in space to a link's RSS variance. The model is an ill-posed inverse problem that is highly sensitive to measurement and modeling noise. No unique solution to the least-squares formulation exists, and regularization must be applied to obtain a solution. In this paper, \textit{Tikhonov} regularization is used, but other common forms of regularization as they apply to RTI are discussed and evaluated in \cite{Wilson09b}.

In Tikhonov least-squares regularization, the optimization for image estimation is formulated as
\begin{equation}
\textbf{x}_{Tik} = \arg \min_\textbf{x} \frac{1}{2}||\textbf{Wx} - \hat{\textbf{s}}||^2 + \alpha || \textbf{Qx} ||^2
\label{eq.tikhonov}
\end{equation} 
where $\textbf{Q}$ is the \textit{Tikhonov matrix} that enforces a solution with certain desired properties, and $\alpha$ is a tunable regularization parameter. Taking the derivative of (\ref{eq.tikhonov}) and setting to zero results in the solution:
\begin{equation}
\textbf{x}_{Tik} = (\textbf{W}^T \textbf{W} + \alpha \textbf{Q}^T \textbf{Q})^{-1} \textbf{W}^T \textbf{s}.
\end{equation} 

Tikhonov regularization provides a simple framework for incorporating desired characteristics into the VRTI reconstruction. If smooth images are desired, a difference matrix approximating the derivative of the image can be used as the Tikhonov matrix. If the image is two dimensional, the regularization should include the difference operations in both the vertical and horizontal directions.  Let $\textbf{D}_x$ be the difference operator for the horizontal direction, and $\textbf{D}_y$ be the difference operator for the verticle direction. Then the Tikhonov regularized least-squares solution is
\begin{eqnarray}
\textbf{x}_{Tik} &=& \Pi \hat{\textbf{s}}. \\
\Pi &=& (\textbf{W}^T \textbf{W} + \alpha ( \textbf{D}_x^T \textbf{D}_x + \textbf{D}_y^T \textbf{D}_y)^{-1} \textbf{W}^T \nonumber
\end{eqnarray}

In summary, the variance of each link is estimated from a recent history of RSS samples and stored in vector $\hat{\textbf{s}}$. The regularized image solution is simply a linear transformation $\Pi$ of this vector $\hat{\textbf{s}}$.

\section{Kalman Filter Tracking}\label{section.kalman}
A radio tomography image in itself does not provide the location coordinates of moving objects. The Kalman filter provides a framework to track such coordinate estimates \cite{Maybeck79}, \cite{Welch95}. The Kalman filter is used extensively to estimate the hidden state of a linear system when measurements of that state have been corrupted by noise. It takes into account the current and previous measurements to generate a more accurate estimate of the system's state than a single instantaneous measurement can. A Kalman filter also has the desirable characteristic that the estimate can be updated with each new measurement, without the need to perform batch measurement collection and processing.

In a location tracking system, such as the one described in this paper, the state to be estimated is made up of the physical coordinates of the object being tracked. The Kalman filter exploits the fact that an object moves through space at a limited speed, smoothing the effects of noise and preventing the tracking from ``jumping.'' In this sense, the filter can be viewed as a form of regularization.

To summarize, the Kalman filter algorithm follows a few important steps:
\begin{enumerate}
\item Predict the next state based on known state transition statistics.
\item Take noisy measurement.
\item Compare prediction and measurement to generate an optimal Kalman gain.
\item Combine predicted and measured state estimates to get improved results.
\item Repeat.
\end{enumerate}

In this work, the objects being tracked are assumed to move as a Brownian process, and measurement noise is assumed to be Gaussian. These assumptions are approximations, but the Kalman filter still provides a useful framework for tracking in an RTI system.  The following variables are used in the tracking filter.
\begin{itemize}
\item $\upsilon^2_m$: the variance of the object's motion process, indicating how fast the object is capable of moving. Larger values enable the filter to track faster moving objects, but also make the estimate noisier.
\item $\upsilon^2_n$: the variance of the measurement noise. Larger values will cause the filter to ``trust'' the statistical predictions over the instantaneous measurements.
\item $\textbf{c}$: a two-element vector containing the Kalman estimate for both x and y coordinates.
\item $\textbf{z}$: a two-element vector containing the instantaneous measurement of the target being tracked.
\item $\bar{\textbf{P}}$: the a priori error covariance matrix.
\item $\textbf{P}$: the a posteriori error covariance matrix.
\item $\textbf{G}$: the Kalman gain.
\end{itemize}

With these assumptions and variables, the Kalman filter algorithm for tracking movements in an RTI system can be described by the following steps.
\begin{enumerate}
\item Initialize $\textbf{c}=(0,0)$ and $\textbf{P}=\textbf{I}_2$, where $\textbf{I}_2$ is the 2x2 identity matrix.
\item Set $\bar{\textbf{P}} = \textbf{P} + \upsilon^2_m \textbf{I}_2$.
\item Set $\textbf{G}= \bar{\textbf{P}}(\bar{\textbf{P}}+ \upsilon^2_n \textbf{I}_2)^{-1}$.
\item Take measurement $\textbf{z}$ equal to the coordinates of the maximum of the VRTI image.
\item Set $\textbf{c} = \textbf{c} + \textbf{G}(\textbf{z}-\textbf{c})$.
\item Set $\textbf{P} = (\textbf{I}_2-\textbf{G}) \bar{\textbf{P}}$.
\item Jump back to step 2 and repeat.
\end{enumerate}

Although this algorithm is designed to track only one object, it could be extended for multiple target tracking. This would entail changing the maximum image coordinate step to detect the instantaneous location of multiple objects within the RTI result. Multiple target tracking with VRTI is a topic for future research.

\section{Experiment}\label{section.experiment}
\subsection{Description and Layout}

This section presents the results of a through-wall tracking experiment utilizing variance-based RTI. A 34-node peer-to-peer network was deployed in an area around a four-wall portion of a typical home. Three of the walls are external, and one is located on the interior of the home. Since this area is an addition to the home, however, the interior wall is made of brick and was an external wall prior to remodeling of the home. Objects like furniture, appliances, and window screens were not removed from the home to ensure that the tracking was functional in a natural environment.

\begin{figure}[ht]
\centering
\includegraphics[scale=.56]{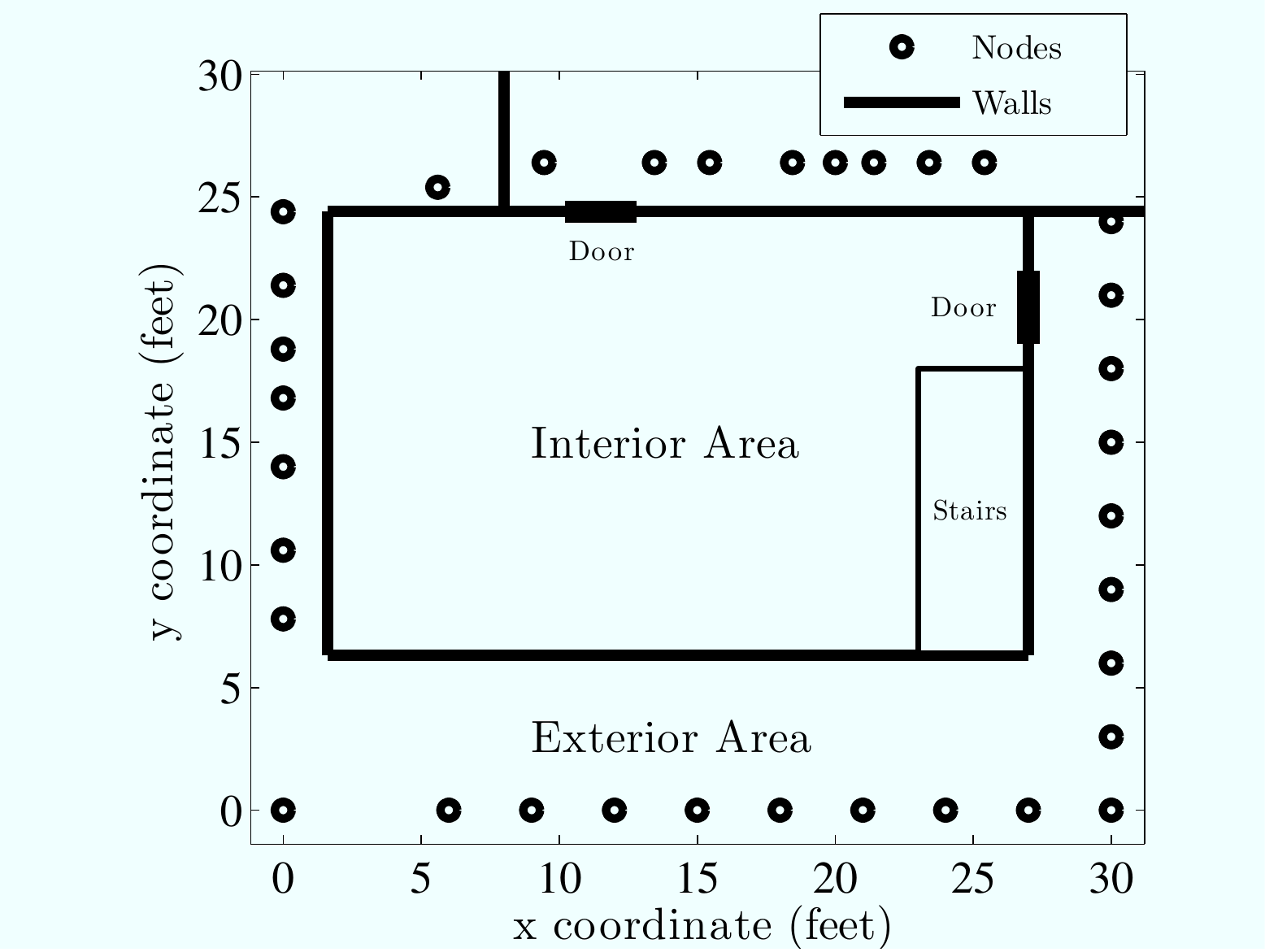}
\caption{The layout of a 34-node variance-based RTI through-wall tracking experiment.}
\label{fig.layout}
\end{figure}

The nodes were placed in a rectangular perimeter, as depicted in Fig. \ref{fig.layout}. It was neither possible, nor necessary, to place the nodes in a uniform spacing due to building and property obstacles. Eight of the nodes were placed on the inside of the building, but on the other side of the brick interior wall. Each radio was placed on a stand to keep them on the same two-dimensional plane at approximately human torso level. 

The nodes utilize the IEEE 802.15.4 protocol, and transmit in the 2.4GHz frequency band. To avoid network transmission collisions, a simple token passing protocol is used.  Each node is assigned an ID number and programmed with a known order of transmission. When a node transmits, each node that receives the transmission examines the sender identification number. The receiving nodes check to see if it is their turn to transmit, and if not, they wait for the next node to transmit. If the next node does not transmit, or the packet is corrupted, a timeout causes each receiver to move to the next node in the schedule so that the cycle is not halted. A base-station node that receives all broadcasts is used to gather signal strength information and save it to a laptop computer for real-time processing.

In all the experimental results in this section, the same set of image reconstruction parameters is used, as shown in Table \ref{table.reconstructionParams}.

\begin{table}[htp]
\centering
\begin{tabular}{|c|c|c|c|}
\hline
\textbf{Parameter} & \textbf{Value} & \textbf{Description} \\ \hline
$\Delta_p$ & 1.5 & Pixel width (ft) \\ \hline
$\lambda$ & .1 & Width parameter of weighting ellipse (ft) \\ \hline
$\delta_c$ & 5 & Pixel correlation (ft) \\ \hline
$\sigma_x^2$ & .5 & Pixel variance $(dB)^2$ \\ \hline
$\alpha$ & 10 & Regularization parameter\\ \hline
$\psi$ & 60 & Variance weighting scale $(dB)^2$\\ \hline
$N_B$ & 136 & Length of RSS buffer \\ \hline
\end{tabular}
\caption{VRTI image reconstruction parameters}
\label{table.reconstructionParams}
\end{table}

Mean-based RTI \cite{Wilson09a} uses the difference in average signal strength to image the attenuation caused by objects in a wireless network. In through-wall imaging, however, the effect of dense walls prevent many of the links from experiencing significant path loss due to a single human obstructing the link. In many cases, multipath fading can cause the mean signal strength to increase when a human obstructs a link.

\begin{figure}[htp]
\centering
\includegraphics[scale=.55]{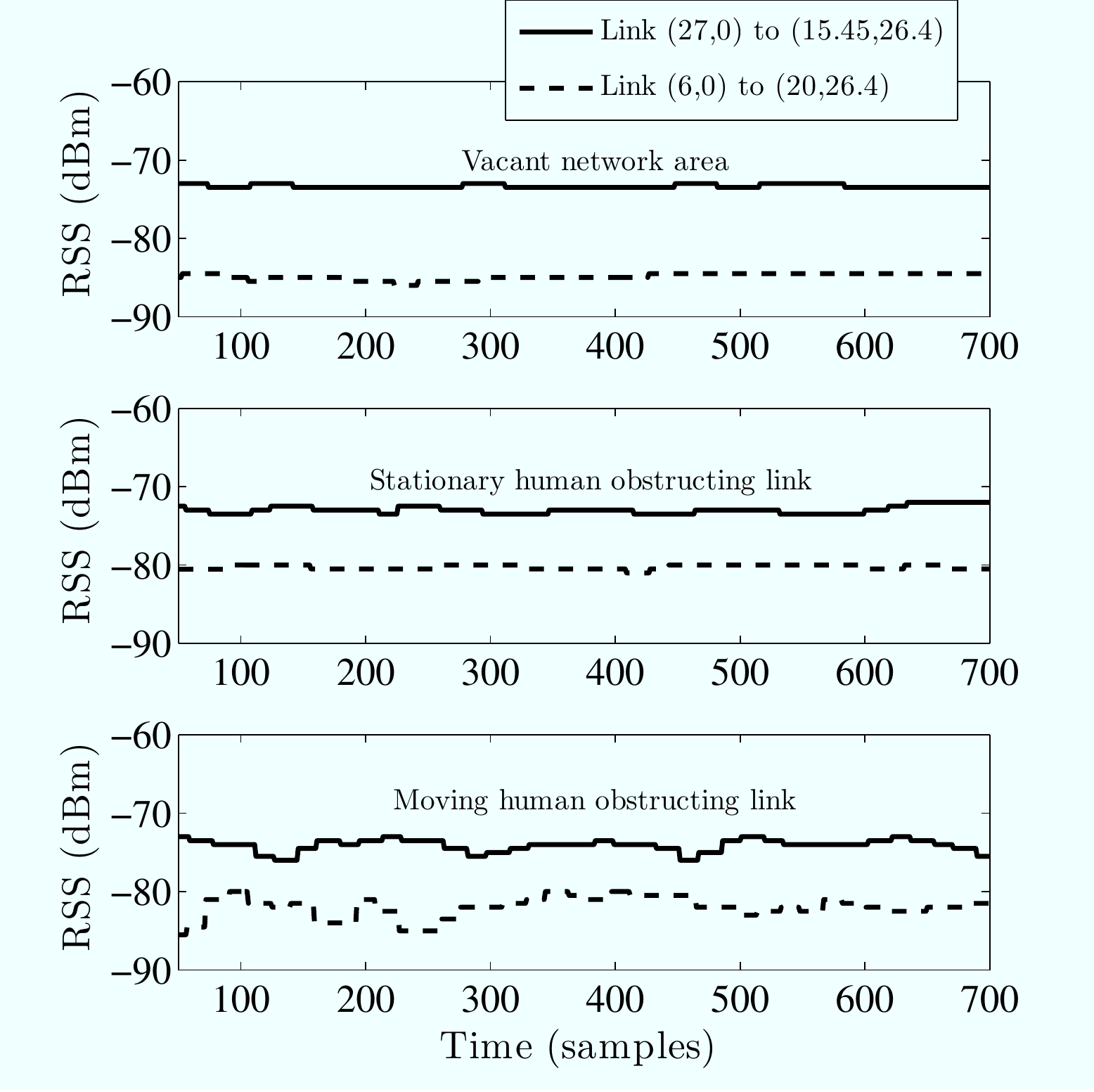}
\caption{RSS measurements for two links in a through-wall wireless network. Comparison of these signals illustrates the advantage of using variance over mean for through-wall imaging human motion.}
\label{fig.timeComparison}
\end{figure}

Variance can be used as an indicator of motion, regardless of the average path loss that occurs due to dense walls and stationary objects within the network. An example of how through-wall links are affected by obstruction is provided in Fig. \ref{fig.timeComparison}.  When a stationary object obstructs the link in a through-wall environment, the change in mean RSS is unpredictable. For example, in Fig. \ref{fig.timeComparison}, one link appears unaffected by the obstruction, while another link's RSS average is raised by approximately 4dB. When an object moves, the variance of the obstructed link's RSS provides a more reliable metric, as seen in the figure.

\subsection{Image Results}
\begin{figure*}[htp]
\centering
\subfigure[Mean-based RTI]{
\label{fig.comparison.RTI}
\includegraphics[scale=.42]{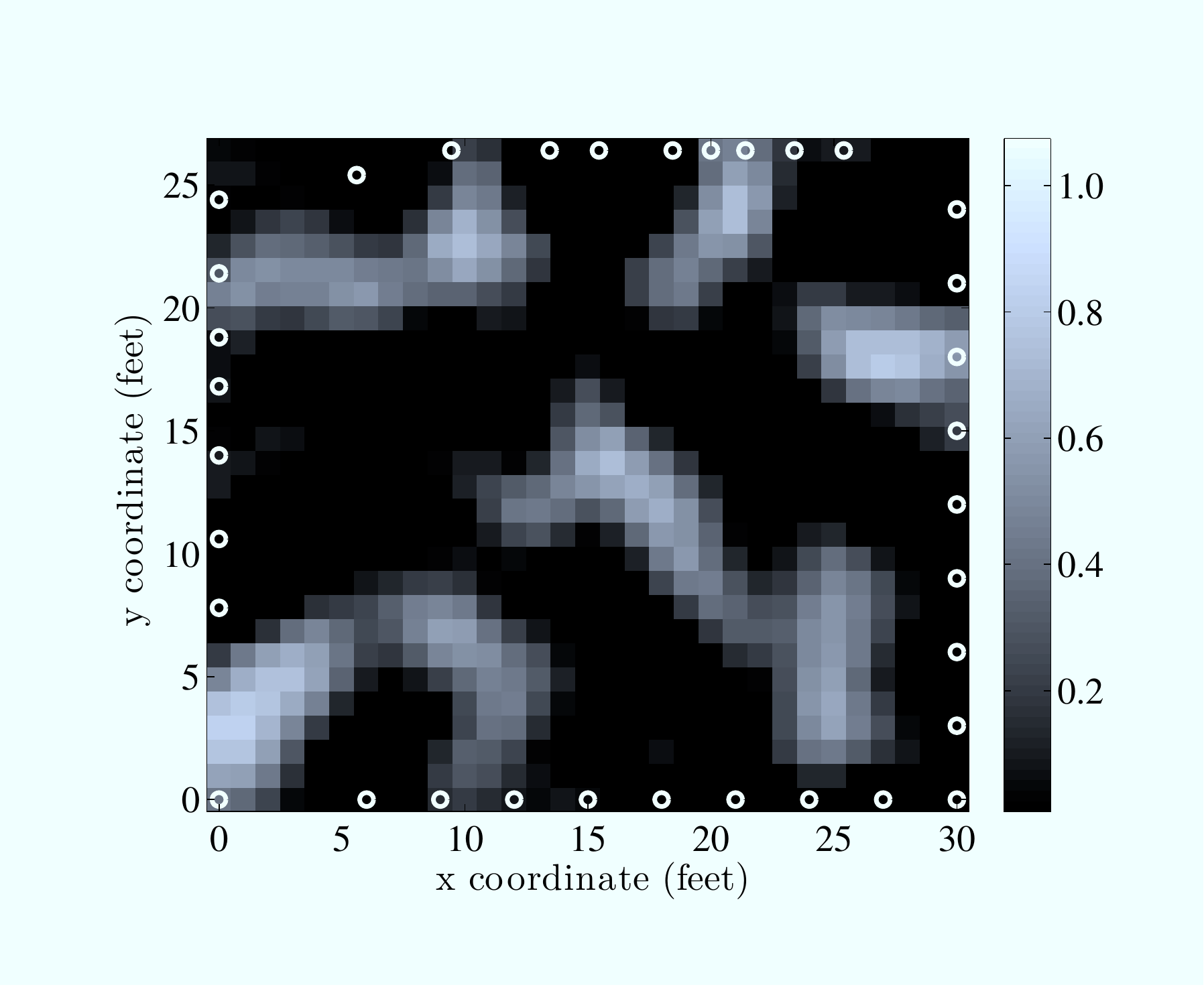}
}
\subfigure[Variance-based RTI]{
\label{fig.comparison.VRTI}
\includegraphics[scale=.42]{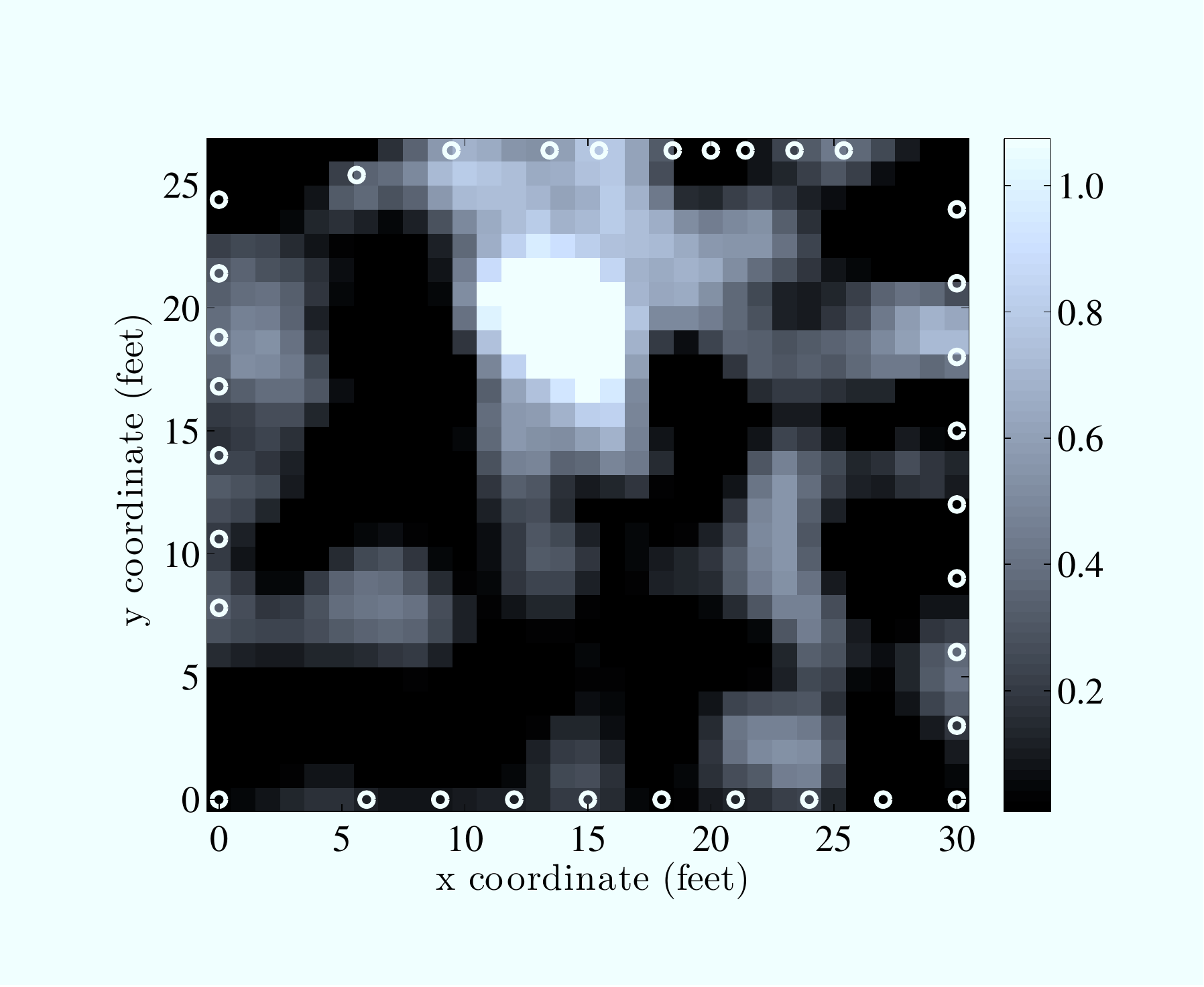}
}
\caption{Comparison of mean and variance-based RTI results for through-wall imaging. The experimenter is moving at point $(17.6,21.3)$ in both of these images.}
\label{fig.comparison}
\end{figure*}
To demonstrate the advantage of using VRTI over mean-based RTI for through-wall motion imaging, two images are presented in Fig. \ref{fig.comparison}. In both images, a human moves randomly, taking small steps around and through the space directly above the coordinate. This is necessary since VRTI images movement, not static changes in attenuation. 

Inspection of Fig. \ref{fig.comparison} shows that VRTI is capable of imaging areas of motion behind walls, while conventional RTI fails to image the change in attenuation.  These results are typical of other location coordinates tested during the experiment.

\subsection{Path Tracking}
In this section, we test Kalman tracking with experimental data. An experimenter moves at a typical walking pace on a pre-defined path at a constant speed. A metronome and uniformly placed markings on the floor help the experimenter to take constant-sized steps at a regular time interval. The experimenter's actual location is interpolated using the start and stop time, and the known marker positions.

\begin{figure*}[htp]
\centering
\subfigure[$\upsilon^2_m = .01$ and $\upsilon^2_n = 5$. Average error $\epsilon = 3.37$ feet]{
\label{fig.path.goodXY}
\includegraphics[scale=.48]{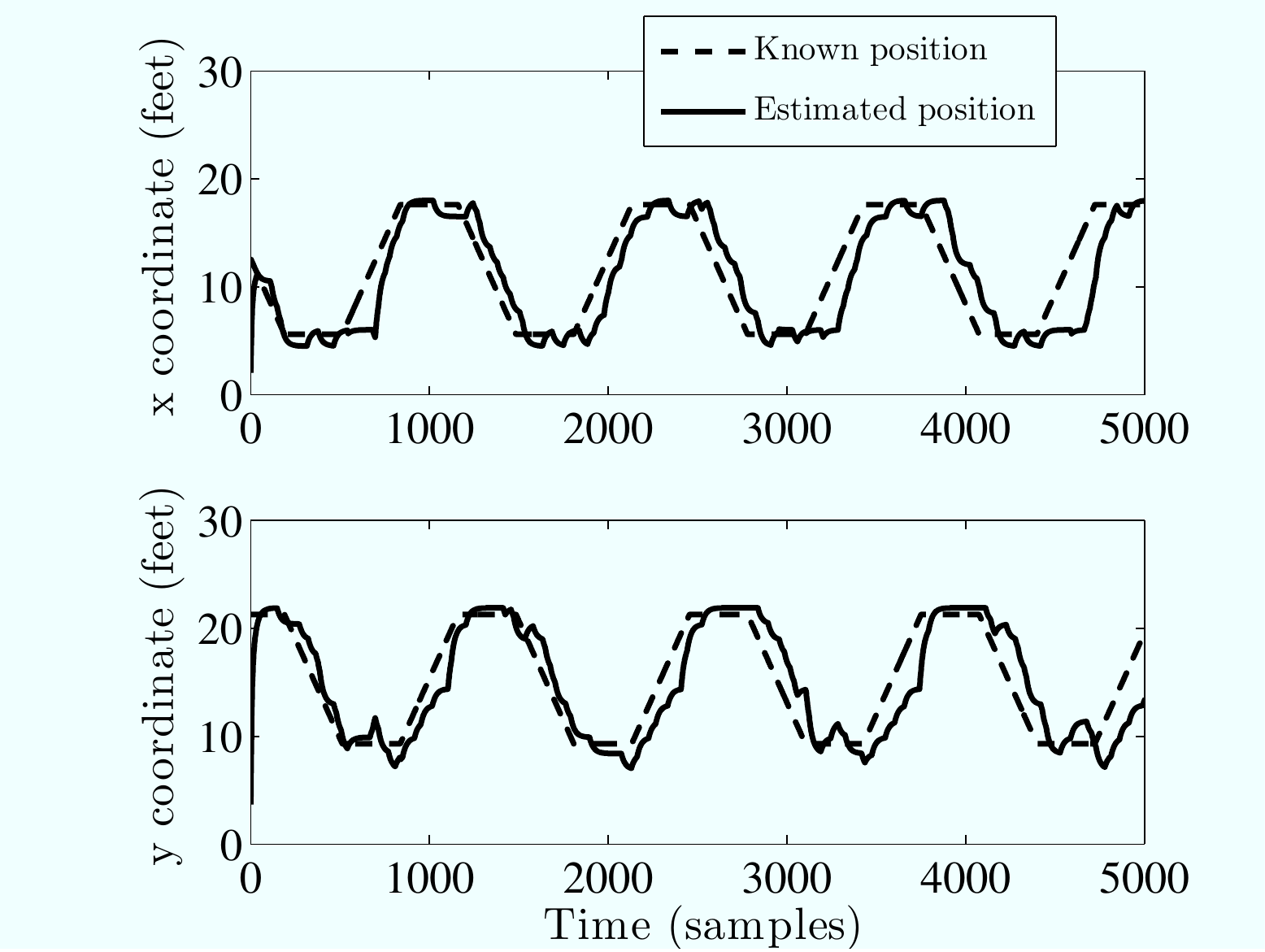}
}
\subfigure[$\upsilon^2_m = .0001$ and $\upsilon^2_n = 5$. Average error $\epsilon = 6.53$ feet]{
\label{fig.path.slowXY}
\includegraphics[scale=.48]{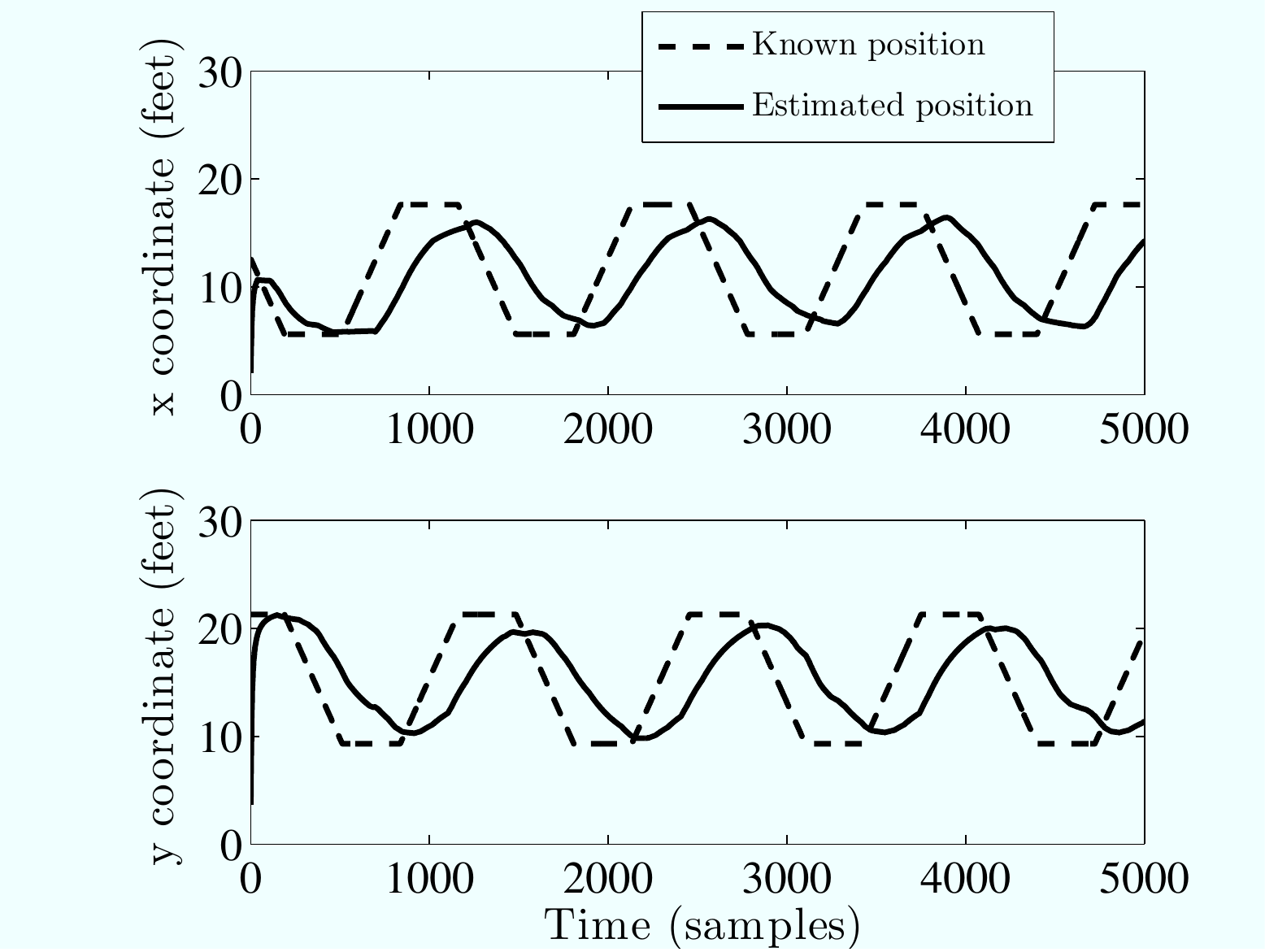}
}
\subfigure[$\upsilon^2_m = .01$ and $\upsilon^2_n = 5$. Average error $\epsilon = 3.37$ feet]{
\label{fig.path.goodPath}
\includegraphics[scale=.48]{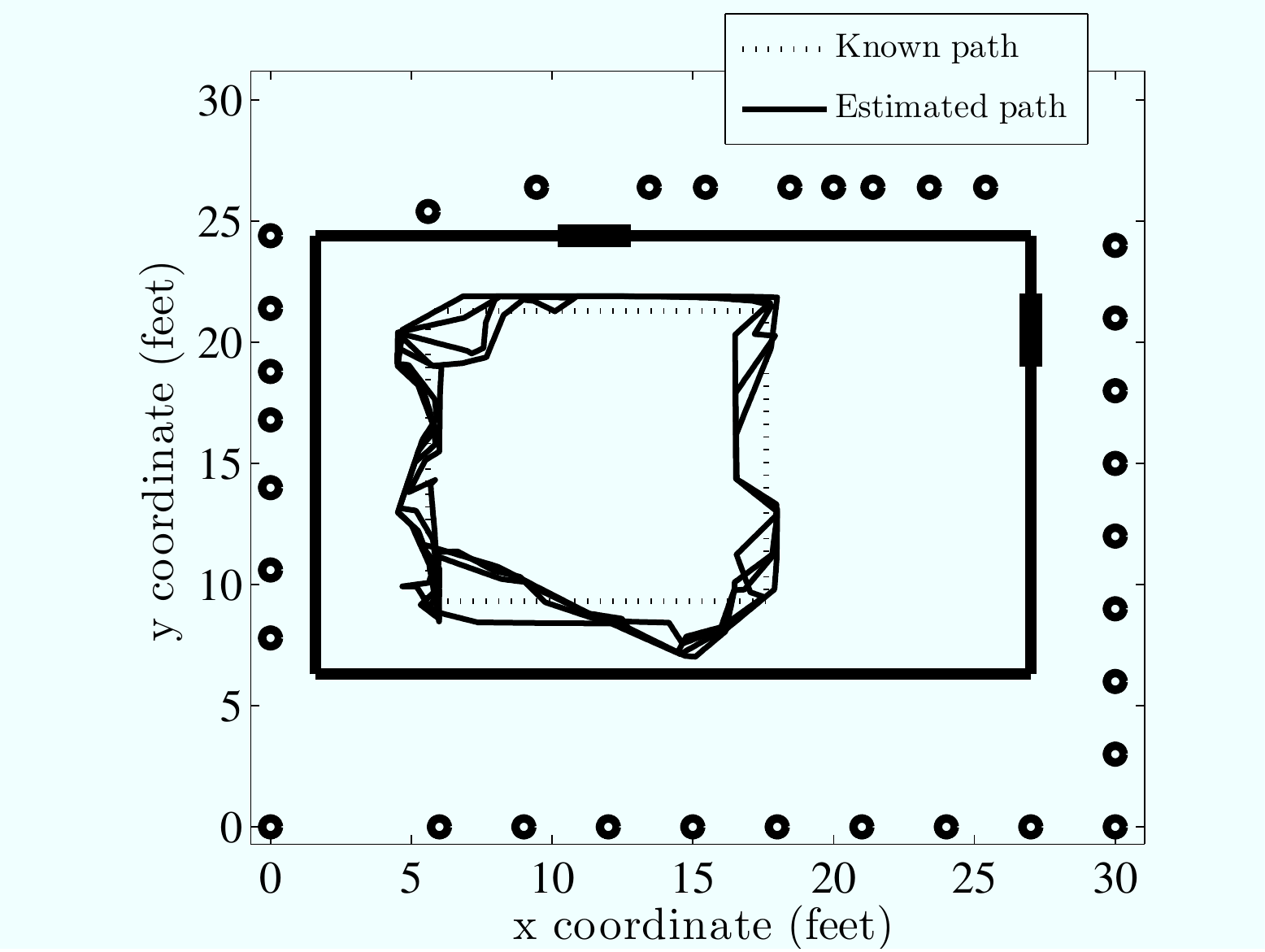}
}
\subfigure[$\upsilon^2_m = .0001$ and $\upsilon^2_n = 5$. Average error $\epsilon = 6.53$ feet]{
\label{fig.path.slowPath}
\includegraphics[scale=.48]{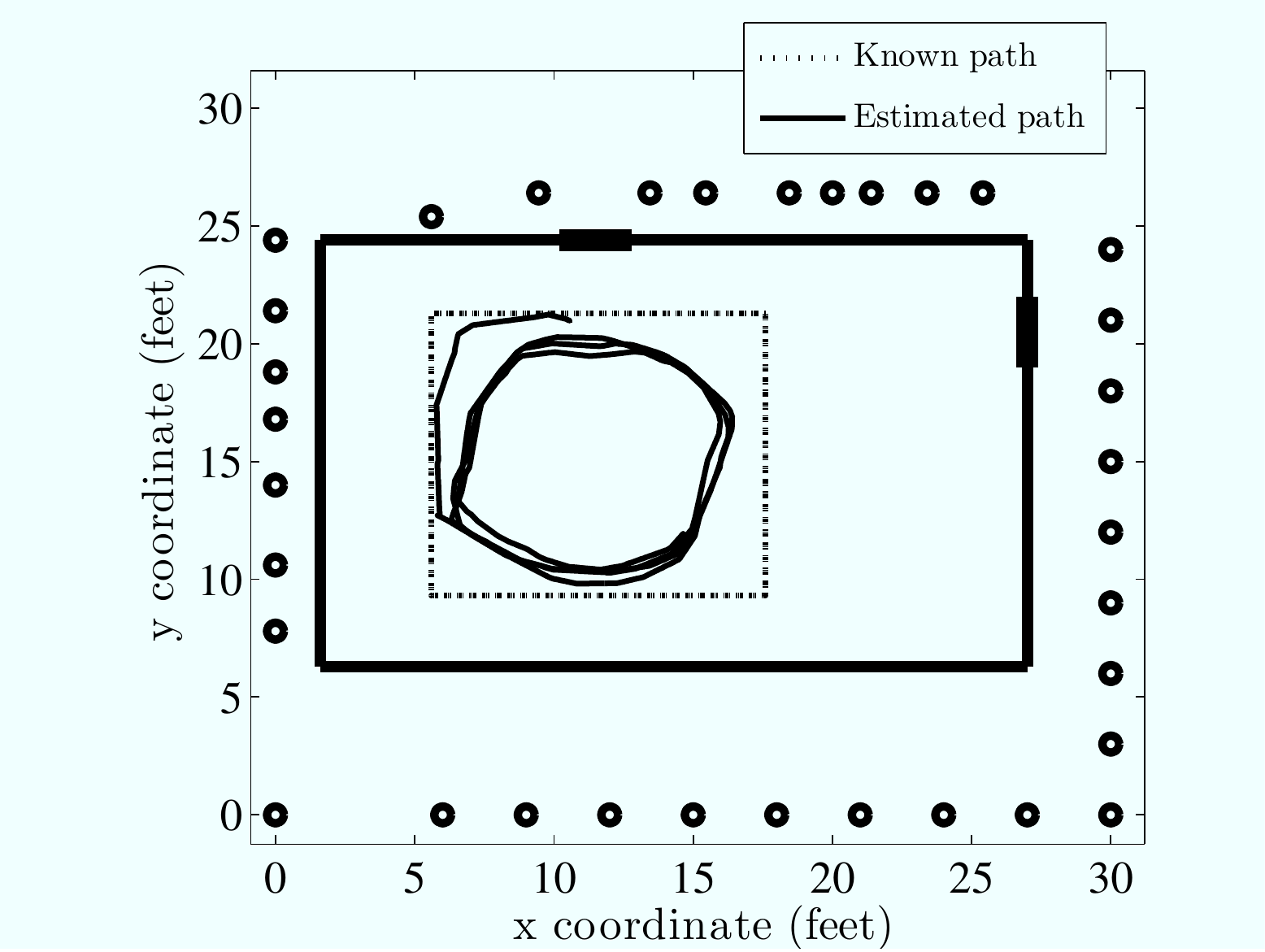}
}
\caption{The location of human movement moving along a known rectangular path is estimated using varying $\upsilon_m$ and constant $\upsilon^2_n = 5$. In (a) and (c) the mobility is set empirically to track objects moving at a few feet per second.  In (b) and (d), the mobility is set too low, causing the tracking filter to lag excessively.}
\label{fig.path}
\end{figure*}

The location of the experimenter is estimated using the Kalman filter described in Section \ref{section.kalman} with tracking parameters presented in Table \ref{table.reconstructionParams}.  Figure \ref{fig.path} plots both the known and estimated location coordinates over time when using two different mobility parameters.

The affect of the Kalman tracking parameters is visually evident in Fig. \ref{fig.path}. When the mobility parameter is set high, the filter is able to track the human with less lag, but the variance of the estimate also increases.  When the mobility parameter is set low, the tracking coordinate severely lags behind the moving object, but estimates a smoother path of motion. 

To quantify the accuracy of the location coordinate estimation, the average error is defined as
\begin{equation} \label{eq.error}
\epsilon = \frac{1}{L} \sum_{k=1}^L \sqrt{(z_x[k] - p_x[k])^2 + (z_y[k] - p_y[k])^2}
\end{equation}
where $L$ is the total number of samples, $z_x[k]$ and $z_y[k]$ are the estimated x and y coordinates at sample time $k$, and $p_x[k]$ and $p_y[k]$ are the actual known coordinates. The average tracking error for $\upsilon^2_m = .01$ and $\upsilon^2_n = 5$ is $3.37$ feet. Other parameters were tested, but none produced more accurate tracking results than these.

It should be noted that a Kalman filter can be designed to estimate the target's velocity, as well as position. This would enable the filter to follow a non-accelerating moving object without a lag. However, when a target changes direction or speed, some transient error would occur while the filter converges to the new speed and direction.

\subsection{Spot Movement}
When estimating the location of a moving object, some amount of tracking lag must occur due to the time it takes to collect measurements from the network and the processing delays.  The lag is also dependent on the mobility parameter $\upsilon_m$ used for tracking.

\begin{figure}
\centering
\includegraphics[scale=.53]{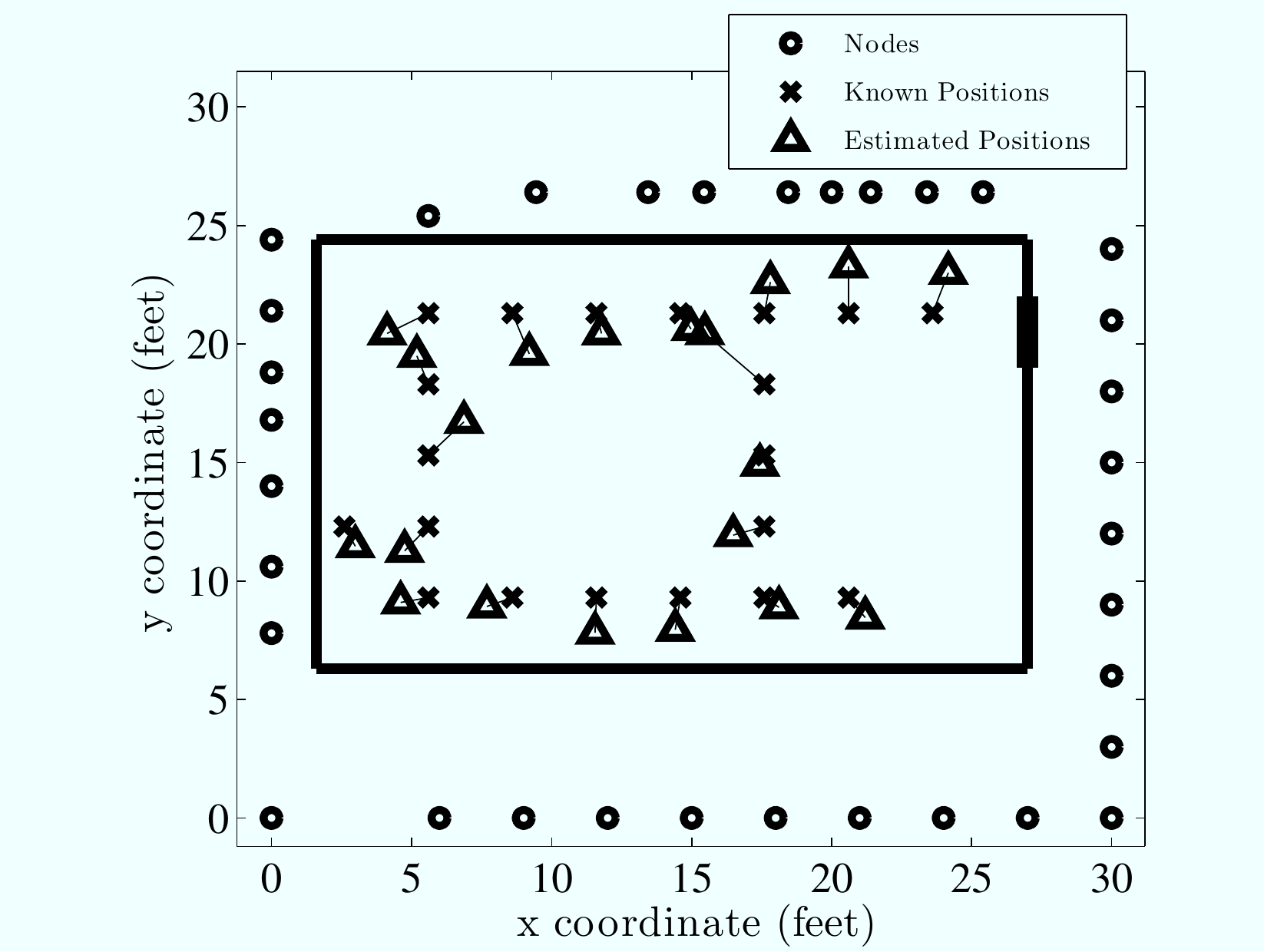}
\caption{The ten-second average locations of human movement over 20 known positions is estimated using VRTI and Kalman filter tracking with parameters $\upsilon^2_m=.01$ and $\upsilon^2_n = 5$. The average error for this experiment $\zeta = 1.46$ feet. }
\label{fig.locations}
\end{figure}

To study the tracking sytem without the effects of time delay, the estimated and known location of a moving human are compared at 20 different coordinates. At each location, the experimenter moves randomly, taking small steps around and through the space directly above the known coordinate. The VRTI tracking system estimates the location of movement and we average the estimates over a duration of ten seconds for each coordinate. The average estimated coordinate is plotted with the known location to generate the results presented in Fig. \ref{fig.locations}.

To quantify the accuracy in this test, the error for each of the 20 known coordinates is averaged.
\begin{equation}
\zeta = \frac{1}{20} \sum_{p=1}^{20} \epsilon_p
\end{equation}
where $\epsilon_p$ is the average error defined by (\ref{eq.error}) for each position $p$. The error for this test for $\upsilon^2_m=.01$ and $\upsilon^2_n = 5$ is $1.46$ feet.

\section{Future Research}\label{section.futureResearch}

Many areas of future research are possible to improve VRTI through-wall tracking technology. First, large and scalable VRTI networks capable of tracking entire homes and buildings need to be explored. This will require advanced wireless networking protocols that can measure the RSS of each link quickly when the number of nodes is high. Perhaps frequency hopping and grouping of nodes will allow a VRTI system to measure each link's RSS while maintaining a low delay in delivering the measurements to a base station.

Advancements on the physical layer modeling will allow VRTI systems to track movement more accurately, and with less nodes.  In this paper, an ellipsoid model is used to relate RSS variance on a link to the locations of movements.  This is certainly an approximation, and future work will require the refinement of the variance weighting model, thus leading to more accurate motion images and coordinate tracking. Other regularization and image estimation techniques may also improve through-wall tracking.

Radio devices could be designed specifically for VRTI tracking applications. The affect of overall node transmission power on imaging performance is an important area to be investigated. Directional and dual-polarized antenna designs would most likely improve images in a through-wall VRTI system. Radio devices capable of sticking to an exterior wall and directively transmitting power into the structure would be extremely useful in emergency deployment and multi-story VRTI.

Finally, localization of nodes plays a significant role in tracking of motion with VRTI networks. In an emergency, rescue or enforcement teams will not have time to survey a location. With automatic node self-localization techniques, the nodes could be thrown or randomly placed around an area and locate themselves without human moderation, saving valuable time.

\section{Conclusion}
Locating interior movement from outside of a building is extremely valuable in emergency situations, enabling police, military forces, and rescue teams to safely locate people prior to entering.  Variance-based radio tomography is a powerful new method for through-wall imaging that can be used to track the coordinates of moving objects. The cost of VRTI hardware is very low in comparison to existing through-wall imaging systems, and a single network is capable of tracking large areas. These features may enable many new applications that are otherwise impractical.

This paper discusses how RSS variance relates to the power contained in multipath components affected by moving objects. The variance of RSS is related to the location of movement relative to node locations, and this paper provides a formulation to estimate a motion image based on variance measurements. The Kalman filter is applied as a mechanism for tracking movement coordinates from image data. A 34-node VRTI experiment is shown to be capable of tracking a moving object through typical home exterior walls with an approximately 3ft average error. An object moving in place can be located with approximately 1.5ft average error.

The experiments presented in this paper demonstrate the theoretical and practical capabilities of VRTI for tracking motion behind walls. Many avenues for future research are presented which may improve image accuracy and enable larger and faster VRTI networks. These future research areas include wireless protocols, antenna design, radio channel modeling, localization, and image reconstruction.

\section*{Acknowledgements}
This material is based upon work supported by the National Science Foundation under the Early Career Faculty Development (CAREER) Grant No. ECCS-0748206. Any opinions, findings, and conclusions or recommendations expressed in this material are those of the authors and do not necessarily reflect the views of the National Science Foundation.

\bibliographystyle{ieeetr}
\bibliography{refs}

\end{document}